\documentclass[10pt]{iopart}

\usepackage{graphicx}
\usepackage[breaklinks=true,colorlinks=true,linkcolor=blue,urlcolor=blue,citecolor=blue]{hyperref}
\usepackage{cite}
\usepackage{amssymb}

\begin{document}

\title[Electron-ion collision spectroscopy at CRYRING@ESR]{Electron-ion collision spectroscopy at the CRYRING@ESR electron cooler}

\author{C~Brandau$^{1,2}$, S~Fuchs$^{1,3}$, V~Hannen$^4$, E~O~Hanu$^{2}$, C~Krantz$^{2}$, M~Lestinsky$^{2}$, M~Looshorn$^{1,3}$, E~B~Menz$^{2,5}$, S~X~Wang$^{1,3}$, and S~Schippers$^{1,3}$}
\address{$^1$I. Physikalisches Institut, Justus-Liebig-Universit\"at Gie\ss{}en, 35392 Giessen, Germany}
\address{$^2$GSI Helmholtzzentrum f\"{u}r Schwerionenforschung GmbH, 64291 Darmstadt, Germany}
\address{$^3$Helmholtz Forschungsakademie Hessen f\"ur FAIR (HFHF), GSI Helmholtzzentrum f\"ur Schwerionenforschung, Campus Gie{\ss}en, 35392 Giessen, Germany}
\address{$^4$Institut f\"ur Kernphysik, Universit\"at M\"unster, 48149 M\"unster, Germany}
\address{$^5$Institut für Kernphysik, Universität zu Köln, 50937 Köln, Germany} 
\ead{stefan.schippers@uni-giessen.de}
\vspace{10pt}
\begin{indented}
\item[]\today
\end{indented}

\begin{abstract}
Electron-ion collision spectroscopy at heavy-ion storage rings aims at precision measurements of resonance features that occur in the cross sections of electron collision processes such as electron-impact ionization of ions or electron-ion recombination. As part of the international FAIR project, the low-energy ion storage ring CRYRING@ESR has been coupled with the heavy-ion accelerators operated by the GSI  Helmholtz Center for Heavy-Ion Research in Darmstadt, Germany. This has created a new opportunity for stringent tests of strong field quantum electrodynamics by electron-ion collision spectroscopy of heavy few-electron ions. The present contribution provides details of the electron-ion collision spectroscopy setup at CRYRING@ESR and of the associated data-analysis procedures along with first results for nonresonant and resonant recombination of berylliumlike lead ions. For nonresonant recombination a recombination rate enhancement factor of 3.5 is found at zero electron-ion collision energy. For resonant recombination there is excellent agreement with recent theoretical results when these are shifted by 340~meV in energy.
\end{abstract}

%
%
\submitto{\CPC}
%
%
\ioptwocol

\section{Introduction}\label{sec:Introduction}

The low-energy heavy-ion storage ring CRYRING \cite{Abrahamsson1993a} was originally located at the Manne-Siegbahn Laboratory in Stockholm, Sweden, where it served, among others, for electron-ion collision experiments with multiply charged atomic ions \cite{Schuch2007a}. The ion species with the highest charge state stored in CRYRING was Pb$^{54+}$ \cite{Lindroth2001}. It was provided by a cryogenic electron-beam ion source, which also gave CRYRING its name. In 2013, CRYRING was transferred to Darmstadt, Germany, as a Swedish in-kind contribution to the international Facility for Antiproton and Ion Research (FAIR). At its new location, the ring was coupled with the heavy-ion accelerator complex of the GSI Helmholtz Center for Heavy-Ion Research consisting of the universal linear accelerator UNILAC, the heavy-ion synchrotron SIS18 and the experimental storage ring ESR and, thus, became CRYRING@ESR. 

The GSI accelerator complex is capable of delivering intense beams of heavy ions with the highest charge states to CRYRING@ESR (e.g.\ Pb$^{82+}$ \cite{Zhu2022}), which facilitates a unique research program in atomic, nuclear and fundamental physics \cite{Lestinsky2016}. Part of the research program addresses electron-ion collision spectroscopy of highly-charged heavy ions, which aims at testing strong-field quantum electrodynamics (QED) with highest precision. Electron-ion collision spectroscopy exploits resonance features that occur in the cross sections of atomic processes such as electron-impact ionization of ions or electron-ion recombination \cite{Mueller2008a}. 

Recently, we have reported first results on resonant electron-ion recombination of berylliumlike Pb$^{78+}$ ions  \cite{Schippers2025}, where the experimental uncertainty was already lower than the corresponding theoretical one \cite{Malyshev2024}, thus constraining second order strong-field QED contributions to $2s\to2p$ core excitation energies. In the present paper, we provide a detailed description of the experimental and data analysis procedures pertaining to electron-ion collision spectroscopy at CRYRING@ESR along with further results for Pb$^{78+}$.

\section{Experimental setup}\label{sec:Setup}

\begin{figure*}
	\centering\includegraphics[width=0.8\textwidth]{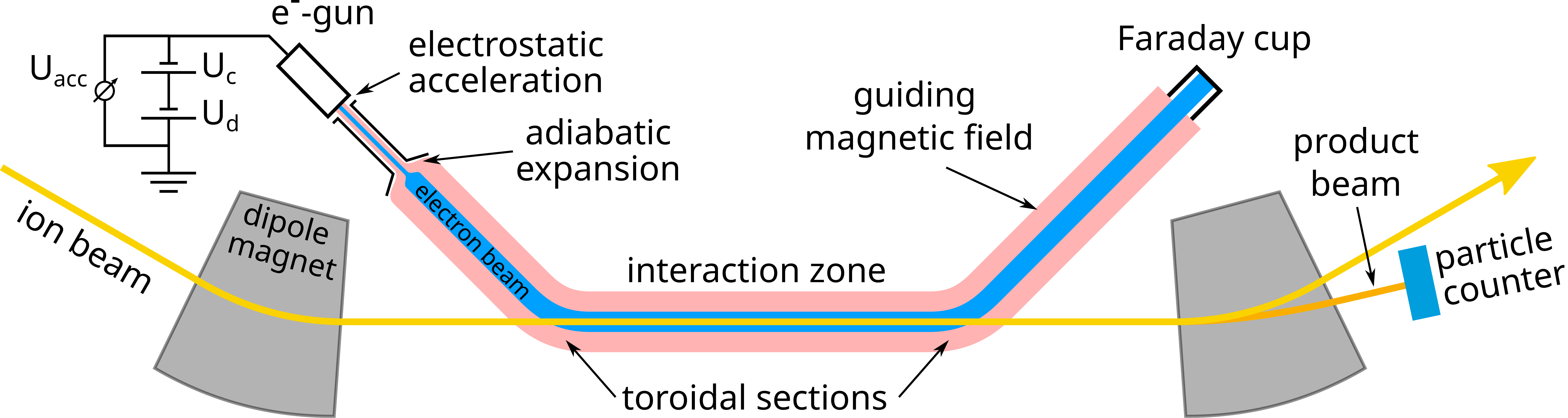}
	\caption{\label{fig:setup}Schematic of the experimental arrangement for electron-ion collision spectroscopy at the CRYRING@ESR electron cooler, which is located in one of the twelve straight sections of the storage ring.}
\end{figure*}

Electron-ion collision spectroscopy at CRYRING@ESR employs the ultra-cold electron beam of its electron cooler \cite{Danared1993a,Danared1994,Krantz2021,Koutsostathis2024} as an electron target for electron-ion recombination. Figure~\ref{fig:setup} sketches the experimental electron-ion merged-beams arrangement. In the electron cooler, the electron beam is magnetically guided onto the ion orbit such that electrons and ions move coaxially in the same direction. Recombined product ions have a lower charge than the stored primary ions and therefore leave the closed orbit in the first dipole magnet downstream of the electron cooler. The recombined ions are counted by an appropriately placed single-particle detector (for details see \cite{Lestinsky2022}). 

Electron cooling forces the velocity $v_i$ of ion beam towards the average electron beam velocity $v_e$ and leads to a decrease of
the ion-beam diameter and its internal velocity spread \cite{Danared1994,Poth1990}. The electron energy, at which the cooling condition, i.e.\ $v_i=v_e$, is realized, is referred to as the cooling energy, $E_\mathrm{cool}$. In electron-ion collision spectroscopy, recombined ions are counted as a function of the electron-ion collision energy. The latter is varied on a millisecond time scale by changing the electron cooler's cathode voltage, as explained in more detail in section~\ref{sec:cc} below.

\begin{figure}
	\centering\includegraphics[width=0.87 \columnwidth]{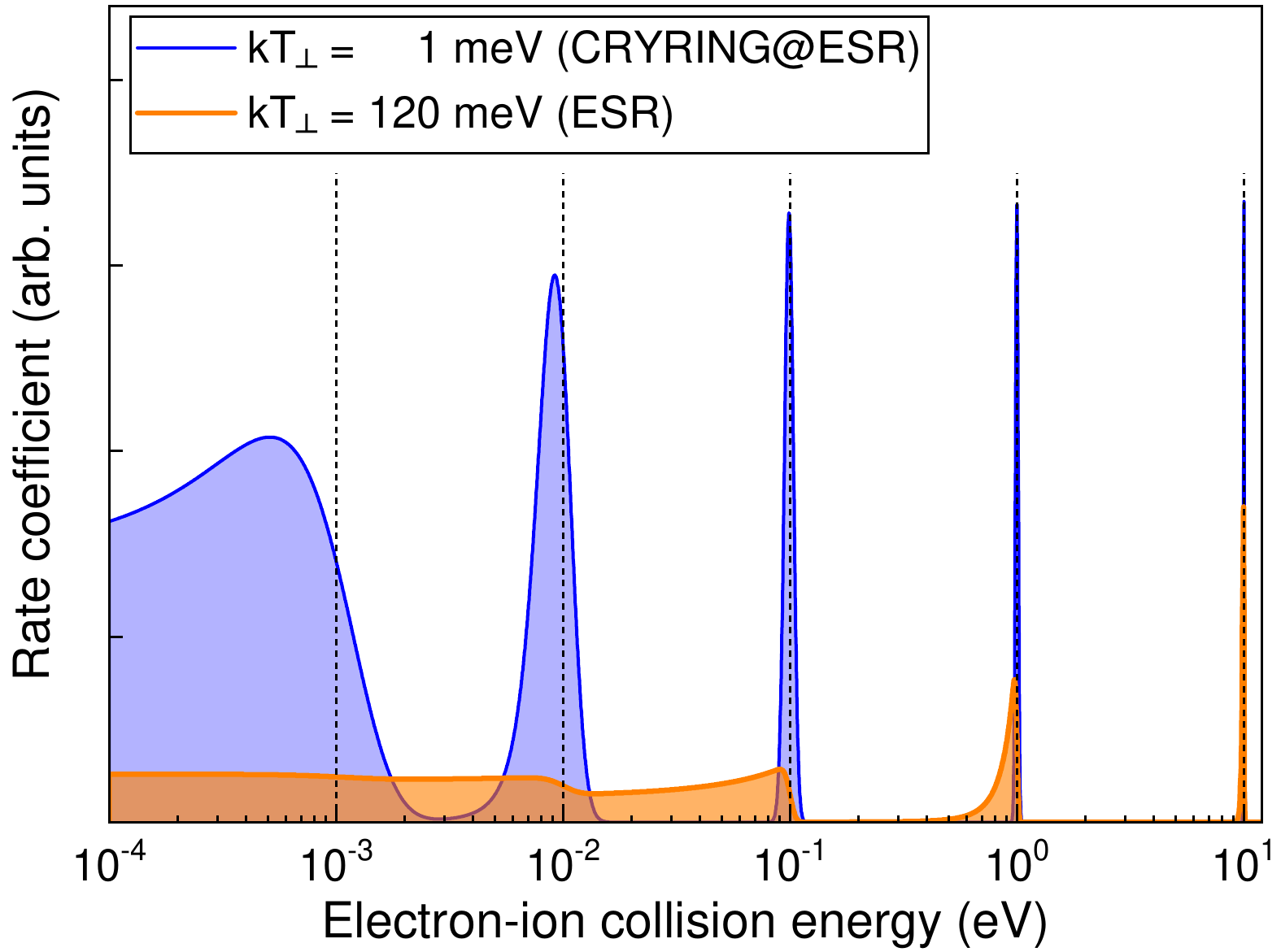}
	\caption{\label{fig:deltares}Synthetic recombination spectra with delta-like resonances at the indicated positions (dashed vertical lines). The shaded curves result from convolutions of the resonances with the cooler electron energy distributions at the heavy-ion storage rings CRYRING@ESR (blue curve, $kT_\perp=1$~meV, $kT_\| = 0.1$~meV) and ESR (orange curve, $kT_\perp=120$~meV, $kT_\| = 0.1$~meV). The immense gain in resolving power when going from ESR to CRYRING@ESR is obvious.}
\end{figure}

Since electron-ion collision spectroscopy aims at an accurate mapping out of recombination resonances (see, e.g., \cite{Lestinsky2008a,Brandau2008a,Bernhardt2015a}) an as-low-as-possible energy spread, $\delta E_\mathrm{cm}$, of the electron beam in the electron-ion center-of-mass frame is of primary concern. The energy spread can be expressed as \cite{Mueller1999c}
\begin{equation}
\delta E_\mathrm{cm} = \sqrt{(kT_\perp\ln2)^2+16E_\mathrm{cm}kT_\|\ln2}.
\end{equation}
{\color{black}where $k$ is Boltzmann's constant and} $T_\|$ and $T_\perp$ are the electron beam's  parallel and transverse temperatures, respectively, with respect to the beam propagation direction. The energy spread increases with increasing electron-ion center-of-mass energy $E_\mathrm{cm}$. Therefore, ions which exhibit recombination resonances close to $E_\mathrm{cm}=0$~eV are particularly suited to being studied by electron-ion collision spectroscopy. In the CRYRING@ESR electron cooler, the transverse temperature is determined by the electron gun's cathode temperature, $T_\mathrm{cath}$, and the beam expansion factor, $\chi=B_\mathrm {cath}/B$, which is realized by passing the electron beam from a region of initially high magnetic guiding field $B_\mathrm{cath}$ generated by a superconducting solenoid to a region of lower field $B$ (figure~\ref{fig:setup}). The maximum value for the beam expansion factor is 100. Due to the acceleration of the electron beam in longitudinal direction, $T_\|$ is usually much lower than $T_\perp$. At CRYRING, $kT_\perp = kT_\mathrm{cath}/\chi = 100\mathrm{~meV}/100 = 1$~meV and $kT_\| = 0.08$~meV can be expected \cite{Lindroth2001}. Figure~\ref{fig:deltares} shows that particularly resonances at energies below about 10~eV benefit immensely from the ultra-cold electron beam at CRYRING@ESR as compared to the conditions at the ESR electron cooler.

\subsection{Cooler control}\label{sec:cc}

\begin{figure}
	\centering\includegraphics[width=0.9\columnwidth]{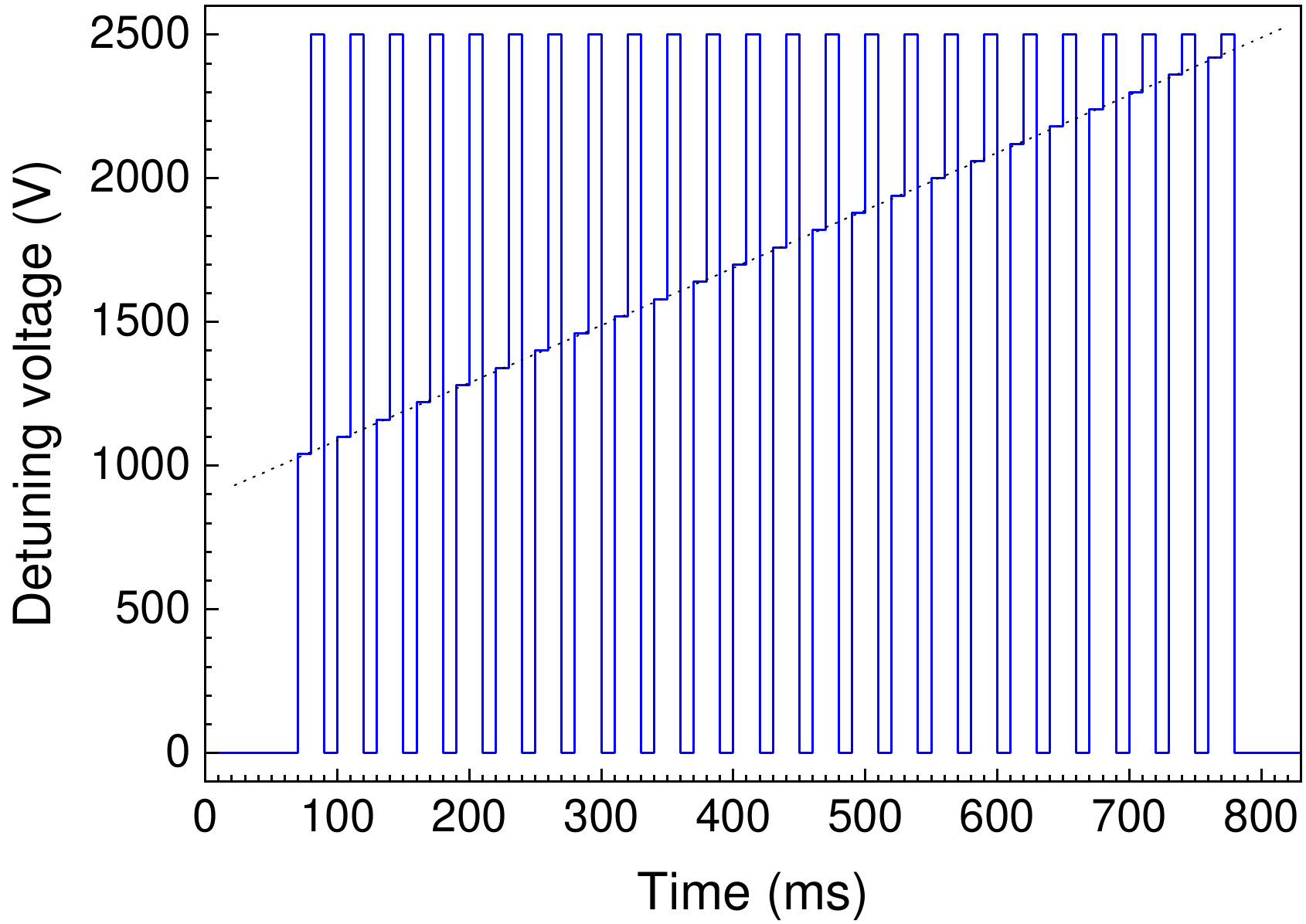}
	\caption{\label{fig:ramp}Example of a voltage ramp for recombination measurements, where the detuning voltage is alternatingly switched between cooling ($U_d^{(c)}=0$~V), measurement ($1000~\textrm{V} \leq U_d^{(m)}\leq 2420$~V), and reference ($U_d^{(r)}=2500$~V). The time duration of each voltage step is 10~ms. The dotted line is to guide the eye along the sequence of measurement voltages. }
\end{figure}

\begin{figure*}
	\includegraphics[width=\linewidth]{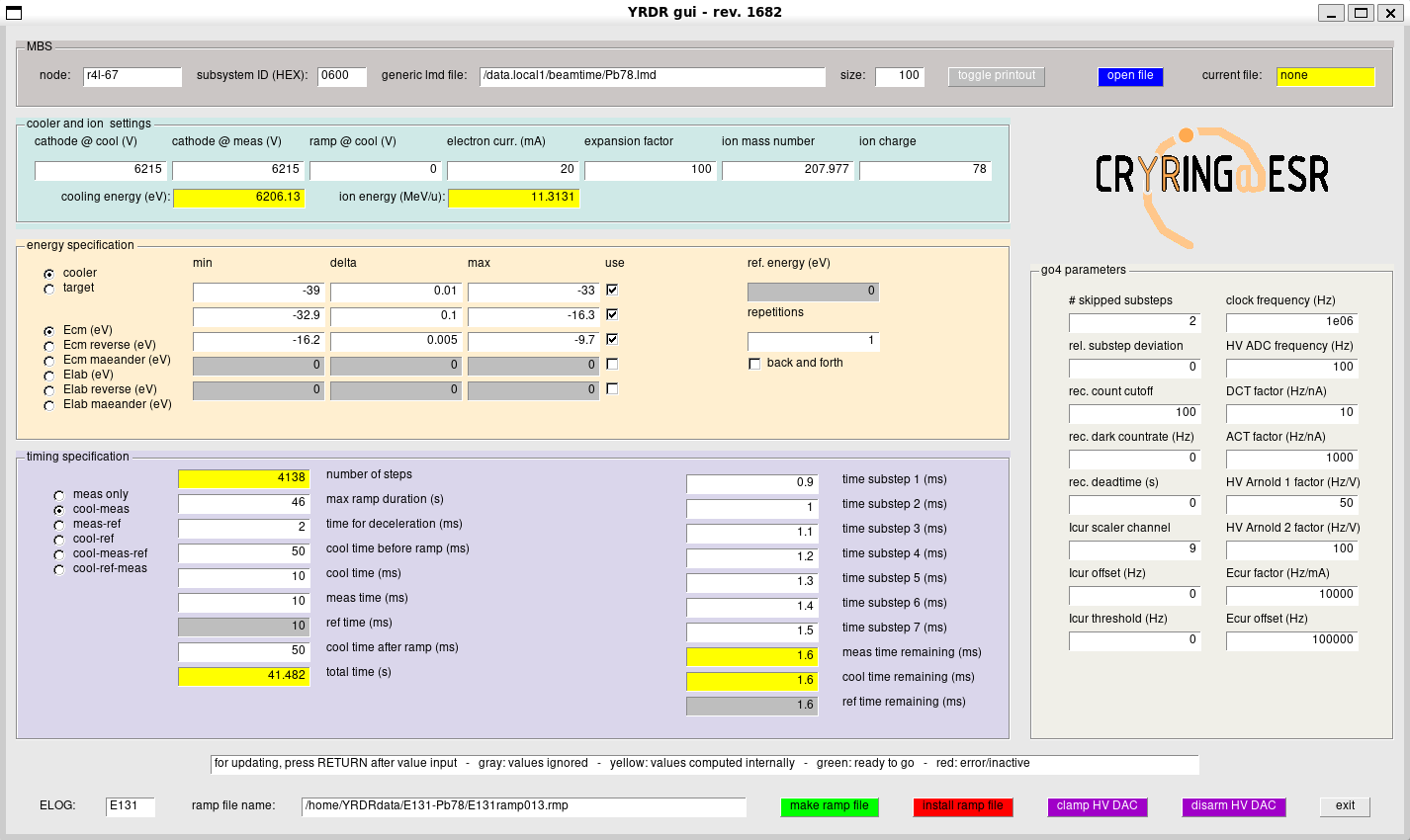}
	\caption{\label{fig:gui}ROOT \cite{Brun1997} based graphical user interface (GUI) for the control of electron-ion recombination measurements at the CRYRING@ESR cooler.   }
\end{figure*}

The electron energy in the laboratory frame is determined by the electron cooler's cathode voltage which corresponds to the electrons' acceleration voltage, $U_\mathrm{acc} = U_c+U_d$, and which is generated by two high-voltage supplies in series (figure~\ref{fig:setup}). A slow high-voltage supply delivers the voltage $U_c$, which is set such that the cooling energy $E_\mathrm{cool}$ is realized. A  fast high-voltage amplifier (Kepco BOP 1000M) provides the detuning voltage $U_d$, which is varied during a measurement while $U_c$ is kept constant. The input voltage of the high-voltage amplifier is generated by a 20-bit DAC (Analog Devices AD5791), which is controlled by a  microcontroller (Teensy 3.6). The latter is connected via USB to a Linux computer (Raspberry Pi 4), which can be accessed via the local network. Before the start of a recombination measurement an ascii file, termed \lq{}ramp file\rq, containing the sequence of voltage values (figure~\ref{fig:ramp}), that will consecutively be set during a voltage ramp, along with the corresponding timing information needs to be uploaded to the microcontroller. A graphical user interface (GUI, figure~\ref{fig:gui}) is available for the generation and upload of the ramp file to the microcontroller. 

\subsection{High-voltage measurement}\label{sec:HV}

Precision collision spectroscopy requires an accurate knowledge of the velocities of the colliding particles. To this end, the acceleration voltage $U_\mathrm{acc}=U_c+U_d$  of the electrons is recorded by high-voltage probes for each of the voltage steps in the voltage ramp. The high-voltage probes must be able to follow the changes of $U_\mathrm{acc}$, which occur on the millisecond time scale. For this purpose two fast high voltage probes have been built by the Giessen and M\"unster groups. 

The Giessen device (known as \lq{}Arnold\rq) was developed more than two decades ago and used already at the ESR storage ring (see e.g.\ \cite{Wang2024}). It is based on a commercial 200-M$\Omega$ voltage divider (HVT-21 by EBG Austria) with a divider ratio of 1000:1. The HVT divider is only 52~mm $\times$ 25.4~mm in size and is mounted in a closed metal cylinder. It was developed to enable a very fast high-voltage measurement with {\color{black}rise times of less than 40~$\mu$s when going from 10\% to 90\% of the requested voltage change and} without any capacitive corrections. This goal is achieved by means of a stainless steel cap that acts as a grading top electrode to minimize stray capacitance, that otherwise in connection with high ohmic value of the resistor chip would lead to long $RC$ times \cite{Kuffel2000}. The output voltage is fed into a temperature-stabilized ($42\pm0.3\,^\circ$C) voltage-to-frequency converter (VFC, Analog Devices AD652BQ) with a maximum output frequency of 1.2~MHz.

The M\"unster fast high-voltage probe (termed ‘FC20’) has been designed and built particularly for its use in collision-spectroscopy experiments at CRYRING@ESR and is a capacitively corrected type. In combination with  a precision multimeter (Keithley DMM7510), which is used for read out, it is about an order of magnitude more precise than the Giessen device (see below).  Further technical details are provided in \cite{Dirkes2020}. Previously, the M\"unster group built an even more precise but slower high-voltage divider (termed \lq{}G35\rq) for DC voltages, which is similar to the KATRIN divider described in \cite{Thuemmler2009} and allows one to measure voltages up to 35~kV with an accuracy of a few ppm ($10^{-6}$). This value includes the uncertainty of the 8.5-digits-voltmeter (Keysight 3458A DVM) that is used for readout. The G35 divider has been calibrated in M\"unster up to 35~kV using a novel absolute calibration method \cite{Rest2020} and later at GSI at 1~kV against a commercial precision divider (Fluke 752A), the 100:1 divider ratio of which has an uncertainty of 0.5~ppm, when properly adjusted.

When the voltage dependence of the scale factors for a voltage range of $\pm 1$~kV around the cooler base voltage was measured, the FC20 divider was stable on the 0.5~ppm level while the Arnold divider exhibited a variation of 95~ppm over a time
interval of several days. When measuring fast voltage changes, as in the actual experiment, the scale factor of the FC20 divider stabilizes within 2~ms inside a $\pm 10$~ppm interval. The FC20 divider therefore allows for an accuracy of $\mathcal{O}(10~{\rm ppm})$ while a measurement with the G35-calibrated Arnold device is expected to be accurate on the 100~ppm level.

\subsection{Data acquisition}

For data acquisition the GSI Multi Branch System (MBS) \cite{MBS2022} is used. The hardware is based on the Versa Module Eurocard-bus (VME). The VME crate houses a central processing unit (CPU) with a Linux operating system that runs the MBS software. The data acquisition is event based, where each event is signalled by a trigger. The triggers are generated either by the accelerator hardware (trigger \lq{}Injection\rq), by the microcontroller that controls the high-voltage amplifier (triggers \lq{}New Ramp\rq\, \lq{}New Voltage\rq, \lq{}End of Ramp\rq), or by a {\color{black} sequencer. The latter provides  up to seven subtriggers for each voltage step. These subtriggers occur after each \lq{}New Voltage\rq\ trigger and are} separated in time by preselected time intervals. The subtrigger timing sequence can be defined in the above mentioned GUI (time substeps 1--7 in figure \ref{fig:gui}).  The timing sequence is transferred to the sequencer via a dedicated Linux computer (Raspberry Pi 4) simultaneously with an upload of a ramp file (section~\ref{sec:cc}).

All trigger signals are fed into a GSI-developed trigger module (GSI TRIVA), which is located in the VME crate. When the trigger module receives a trigger it passes a trigger-characteristic bit pattern to the MBS and initiates a readout of the VME scalers (GSI VULOM), which collect the data from the various measurement devices (detectors, volt meters, amp\`ere meters etc.), most of which are interfaced by VFCs. Another module (GSI VETAR) provides a \lq{}White Rabbit\rq\  \cite{Moreira2009} time stamp for each event, which can be used in the offline analysis for synchronizing data streams from different MBS systems with sub-nanosecond accuracy.  

The data packages, that are generated by each event, consist of a header, a time stamp, an identification of the trigger, and the contents of the scaler channels. The data packages are written consecutively to a file that is linked to the MBS output data stream. During an experiment, this data stream is also directed to an online data-analysis code which visualizes the measured data and derived quantities such as the measured recombination rate coefficient (section \ref{sec:mbrc}). This code is embedded in the GSI software Go4 \cite{Go42024}, which provides convenient access to the MBS data structures. The same code is also used for the refined processing of the data files in the later offline data analysis.

\section{Data analysis} 

\subsection{Electron-ion collision energy}\label{sec:ecm}

\begin{figure}
	\includegraphics[width=\linewidth]{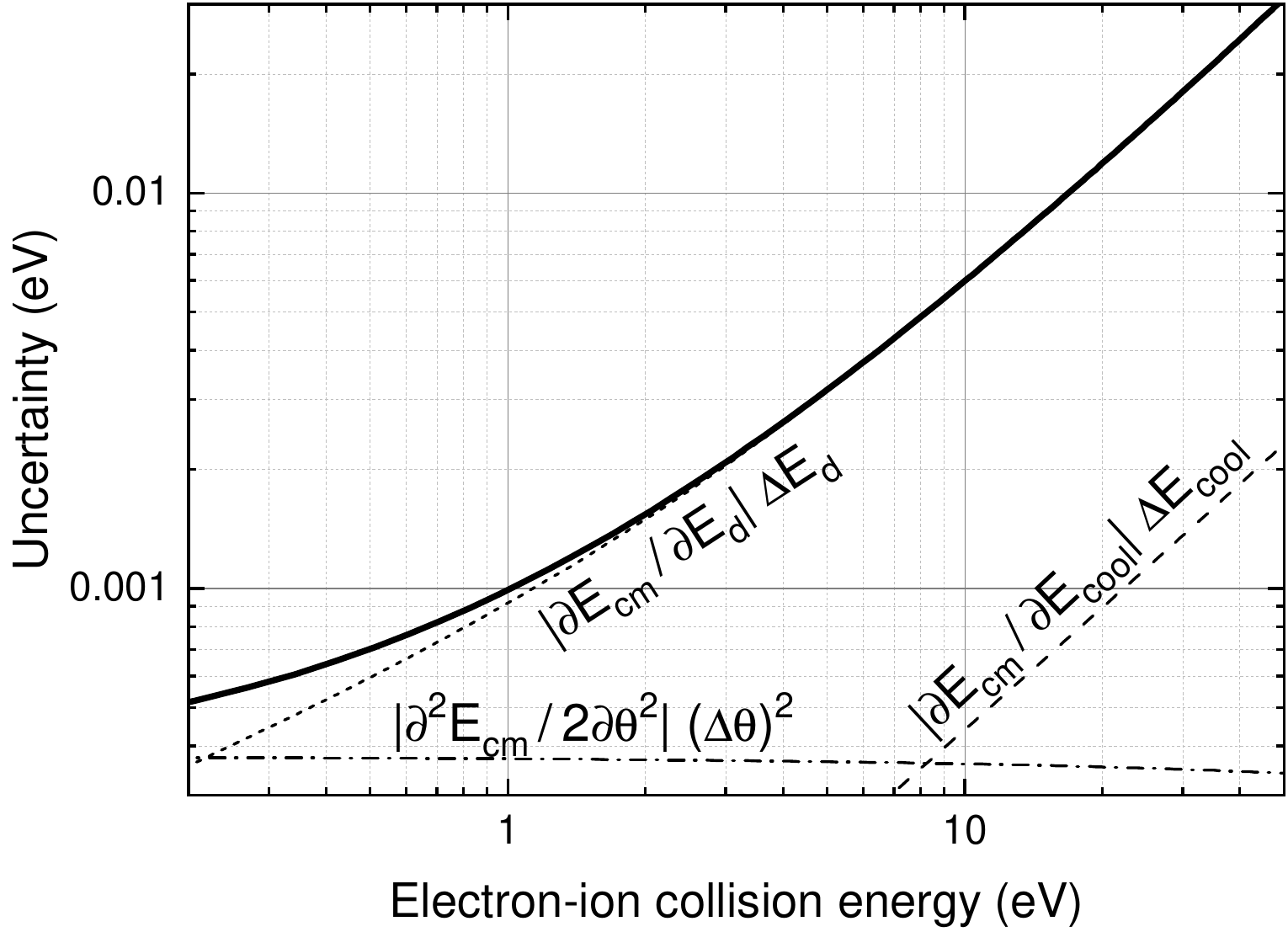}
	\caption{\label{fig:error}Uncertainty of the experimental energy scale and individual contributions according to equation~\eref{eq:dEcm} for $E_\mathrm{cool}=6000$~eV, $m_i = 238$~u, $\Delta E_d/E_d=10^{-5}$, $\Delta E_\mathrm{cool} = 0.25$~eV, $\Delta \theta = 0.25$~mrad. }
\end{figure}

The essential aspect of electron-ion collision spectros\-copy at CRYRING@ESR is an as-accurate-as-possible deter\-mination of the electron-ion collision energy, which requires a careful analysis of all associated uncertainties. The electron-ion collision energy in the center-of-mass frame can be expressed in terms of the laboratory electron and ion velocities $v_e$ and $v_i$, respectively, and the angle $\theta$  between the two interacting beams as \cite{Wang2024}
\begin{equation}
	E_{\mathrm{cm}} = m_ic^{2} (1 + \mu) \left[ \sqrt{1 + \frac{2\mu(\gamma_\mathrm{rel} - 1)}{(1 + \mu)^{2}} } - 1 \right]\label{eq:Ecm}
\end{equation}
where  $c$ is the speed of light in vacuum, $\mu=m_e/m_i\ll1$ is the ratio of the electron rest mass $m_e$  and the ion rest mass $m_i$,  and
\begin{equation}\label{eq:gammarel}
\gamma_\mathrm{rel} = \gamma_{e}\gamma_{i}\left(1 - \beta_e\beta_i\cos\theta\right)
\end{equation}
with  $\beta_e =v_e/c$, $\beta_i=v_i/c$, $\gamma_e=(1-\beta_e^2)^{-1/2}$, and $\gamma_i=(1-\beta_i^2)^{-1/2}$. At the electron-cooling condition, i.e., for $\theta=0$ and $v_i=v_e$, equations~\eref{eq:gammarel} and \eref{eq:Ecm} yield  $\gamma_\mathrm{rel} = 1$ and $E_\mathrm{cm}=0$, respectively,  as expected. The associated electron energy is the cooling energy, $E_\mathrm{cool}$, which is related to $v_i$ via
$\gamma_i = [1-(v_i/c)^2]^{-1/2} = 1 + E_\mathrm{cool}/(m_ec^2)$. When the electron energy is detuned from the cooling energy by an amount $E_d$ this results in $E_e=E_\mathrm{cool}+E_d$ and $\gamma_e = 1 + E_e/(m_ec^2)$. 
The electron-ion collision energy, thus, depends on $E_\mathrm{cool}$, $E_d$, and $\theta$. Consequently its uncertainty can be estimated from the uncertainties $\Delta E_\mathrm{cool}$, $\Delta E_d$, and $\Delta \theta$ as
 \begin{eqnarray}\label{eq:dEcm}
 (\Delta E_\mathrm{cm})^2 &=& \left\vert\frac{\partial E_\mathrm{cm}}{\partial E_\mathrm{cool}}\right\vert^2_{\theta=0}(\Delta E_\mathrm{cool})^2 +\\ 
 && \left\vert\frac{\partial E_\mathrm{cm}}{\partial E_d}\right\vert^2_{\theta=0}(\Delta E_d)^2 + \left\vert\frac{1}{2}\frac{\partial^2 E_\mathrm{cm}}{\partial \theta^2}\right\vert^2_{\theta=0}(\Delta \theta)^4\nonumber\\ &=& \frac{1}{1+2\mu\gamma_\mathrm{rel}+\mu^2} \times\nonumber\\ 
 && \left\{\left\vert\gamma_i\left(1-\frac{\beta_i}{\beta_e}\right)+ \gamma_e\left(1-\frac{\beta_e}{\beta_i}\right)\right\vert^2(\Delta E_\mathrm{cool})^2\right. \nonumber\\
 &&+ \left\vert\gamma_i \left(1-\frac{\beta_i}{\beta_e}\right)\right\vert^2(\Delta E_d)^2\nonumber\\
 &&+ \left.\left\vert \frac{\gamma_e\gamma_i\beta_e\beta_i}{2}\right\vert^2\left[m_ec^2(\Delta \theta)^2\right]^2\right\}.\nonumber
\end{eqnarray}
Since $\partial E_\mathrm{cm}/\partial\theta \propto\sin\theta$ vanishes for $\theta=0$ the second-order term of the Taylor expansion is used for accounting for $\Delta \theta$. Figure~\ref{fig:error} shows an estimate for $\Delta E_\mathrm{cm}$ based on reasonable assumptions for the underlying systematic uncertainties \cite{Schippers2025}. At electron-ion collision energies below 1~eV, $\Delta E_\mathrm{cm}$ is smaller than 1~meV. 

\subsection{Electron-beam space charge}

The electron energy is not only determined by the voltages $U_c$ and $U_d$ of the cathode and detuning voltage supplies, respectively, but also by the space-charge potential, $U_\mathrm{sc}$, of the electron beam, i.e.,
\begin{equation}\label{eq:Usc}
E_e = E_\mathrm{cool}+E_d = -e(U_c+U_d) + eU_\mathrm{sc}.
\end{equation}
$U_\mathrm{sc}$ depends on the radial position in the cylindrical electron beam with radius $r_e = r_c\sqrt{\chi}$, where  $r_c= 0.2$~cm is the radius of the electron coolers' cathode. $U_\mathrm{sc}$ reaches its minimum on the beam axis where it is given as \cite{Boehm2001b} 
\begin{equation}
U_\mathrm{sc} = -\frac{1}{4\pi\varepsilon_0}\left(1+2\ln\frac{r_t}{r_e}\right)\frac{I_e}{v_e},
\end{equation}
where $r_t=5.0$~cm is the radius of the vacuum tube surrounding the electron beam\footnote{In October 2023, i.e.\ after the beamtime with Pb$^{78+}$ ions, a narrower drift tube with $r_t=3.8$~cm was installed inside the vacuum tube of the electron cooler \cite{Koutsostathis2024}.}  and $\varepsilon_0$ is the vacuum electric permittivity. Note, that $U_c+U_d$ and $U_\mathrm{sc}$ are negative and that the space-charge potential thus lowers $E_e$.  Since the space-charge potential via $v_e$ depends on $E_e$ itself, it must be calculated iteratively. Usually, only very few iterations are required for reaching convergence. As explained in \cite{Schippers2025}, the space-charge potential contributes to $E_\mathrm{cool}$ and $E_d$ (equation~\ref{eq:Usc}) and, consequently, its uncertainty to $\Delta E_\mathrm{cool}$ and $\Delta E_d$. This has been taken into account in the error estimate displayed in figure~\ref{fig:error} (see also \cite{Schippers2025}).

\subsection{Merged-beams rate coefficient}\label{sec:mbrc}

The measured merged-beams recombination rate coefficient $\alpha_\mathrm{mb}$ is put on an absolute scale (with typically $\pm$14\% systematic uncertainty \cite{Boehm2001b}) by normalizing the background-subtracted detector count rate $R$ on the number 
$N_i$  of stored ions and on the electron-density $n_e$ using \cite{Wang2024}
\begin{equation}\label{eq:alpha}
	\alpha_\mathrm{mb}(E_\mathrm{cm}) =  \frac{R(E_\mathrm{cm})\,C/L}{[1-\beta_e(E_\mathrm{cm})\beta_i] N_i n_e(E_\mathrm{cm}) \eta}, 
\end{equation}
where $C = 54.178$~m is the ring circumference, $L= 0.9$~m is the length of the electron cooler's straight section, and $\eta$ denotes the detection efficiency of the single particle detector. The number of ions can be calculated from the ion current $I_i$ as 
\begin{equation}\label{eq:Ni}
N_i=\frac{I_i}{ev_i}C,
\end{equation}
where $e$ denotes the elementary charge. The electron density is obtained from the electron current $I_e$ and the electron-beam cross section $A=\pi r_e^2$  as 
\begin{equation}\label{eq:ne}
n_e(E_\mathrm{cm}) = \frac{I_e}{ev_e(E_\mathrm{cm}) A}.
\end{equation}

\subsection{Monte-Carlo convolution of theoretical cross sections}\label{sec:mc}

\begin{figure}
	\includegraphics[width=\linewidth]{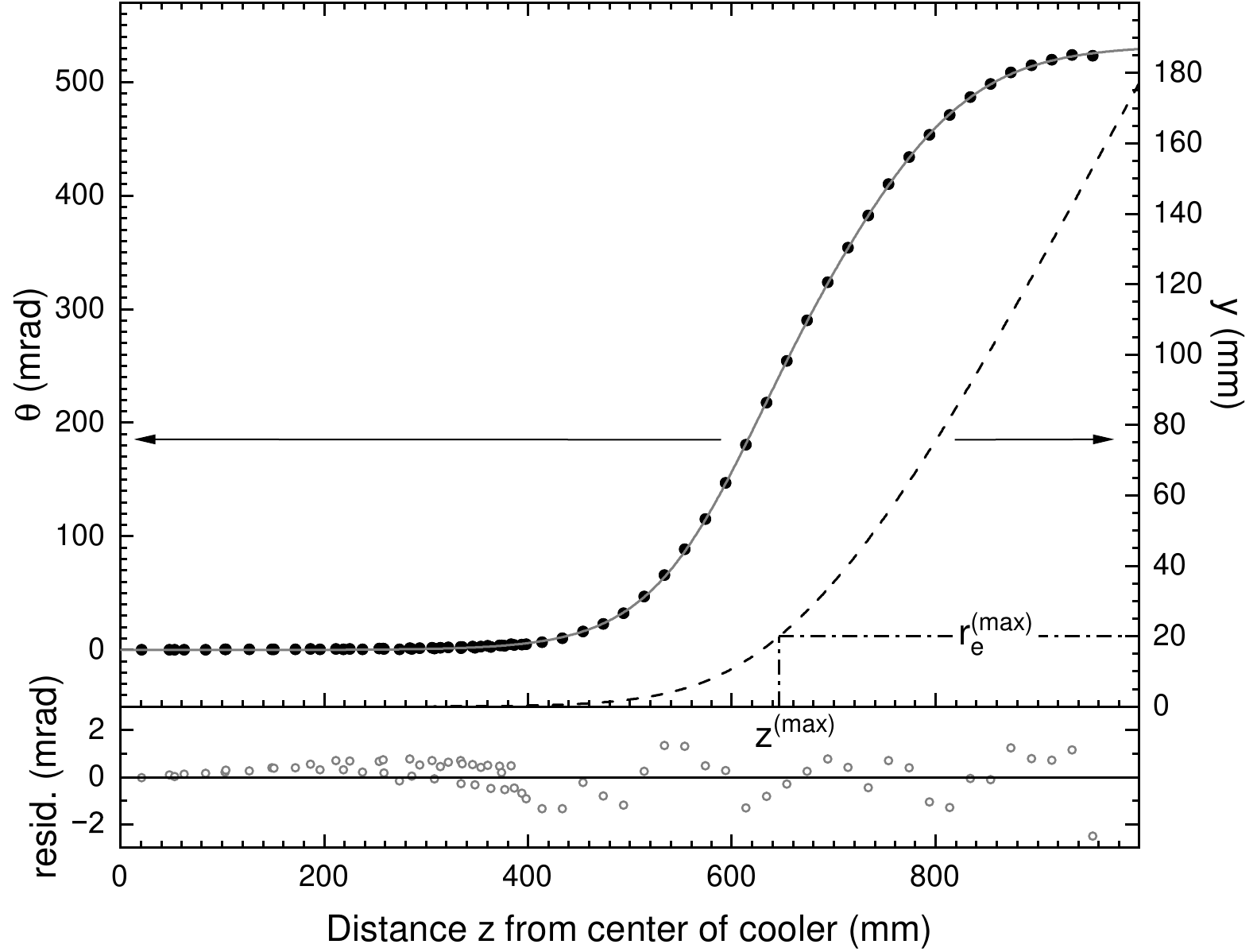}
	\caption{\label{fig:angle}Upper panel: Measured angle of the magnetic field (full symbols, left scale) in the CRYRING electron cooler with respect to the cooler axis and as function of distance $z$ from the cooler center at $z=0$. The angle is essentially zero in the straight section of the cooler and increases in its toroidal sections (figure~\ref{fig:setup}). The full line results from a fit of equation \eref{eq:theta} to the measured data points. The fit parameters are listed in \tref{tab:angle}. The dashed line represents the vertical offset $y(z)$ (right scale) as calculated from equation \eref{eq:y}. The deflection that corresponds to the maximum electron-beam radius ($r_e^\mathrm{(max)}=r_c\sqrt{\chi}=10r_c=20$~mm for the maximum expansion factor $\chi=100$) occurs at z$^\mathrm{(max)} = 646$~mm. Lower panel: Residuals of the fit (open symbols). }
\end{figure}

For the comparison of the measured merged-beams rate coefficients with theoretical cross sections the latter need to be convolved with the experimental electron energy distribution. This has been discussed comprehensively in the appendix of \cite{Wang2024}. The electron energy distribution can be represented analytically as a flattened Maxwellian, which is characterized by the electron beam temperatures $T_\|$ and $T_\perp$. However, a convolution with this distribution function does not account for effects of non-zero angles between electron and ion beam which occur in the merging and demerging regions of the electron cooler, where the electrons follow the magnetic field lines of the respective toroidal magnets. This and other experimental effects (such as the almost negligible energy spread of the ion beam) can be conveniently accounted for by a Monte-Carlo convolution as has been described in \cite{Wang2024} for electron-ion collision spectroscopy at the ESR electron cooler.

The dependence of the angle on the position in the cooler has been obtained from measurements of the magnetic field inside the electron cooler performed during commissioning in Stockholm. We combined high-precision measurements of the field homogeneity near the cooler center, as reported in  \cite{Danared1993a}, with lower-precision vertical field maps of the toroidal sections \cite{Danaredpriv}.  The result is depicted by the full symbols in figure~\ref{fig:angle}. For convenient use in the Monte-Carlo convolution the data points have been fitted by the function
\begin{eqnarray}\label{eq:theta}
\theta(z) &=& \tilde{\theta}(|z|)-\tilde{\theta}(0)\\
\textrm{with} &&\tilde{\theta}(z) = \frac{A_1}{1+\exp\left(\frac{\displaystyle z_1-z}{\displaystyle k_1}\right)}+\frac{A_2}{1+\exp\left(\frac{\displaystyle z_2-z}{\displaystyle k_2}\right)}\nonumber
\end{eqnarray}
The fit result is the full line in \fref{fig:angle}. The residuals of the fit are shown in the lower panel of figure~\ref{fig:angle}. The parameter values that were obtained from the fit are listed in \tref{tab:angle}. 

\begin{table}
\caption{\label{tab:angle}Parameter values obtained from the fit (\fref{fig:angle}) of equation~\eref{eq:theta} to the measured magnetic field in the CRYRING@ESR electron cooler.}
\begin{indented}
	\item[]	\begin{tabular}{cccc}
\br
i & $A_i$ (mrad) & $z_i$ (mm) & $k_i$ (mm) \\
\mr 
1 &307.03794   & 609.73215   &  52.11124   \\
2 & 225.45754   & 745.68523   &  59.52064\\
\br
	\end{tabular}
\end{indented}
\end{table}

Because of the deflection of the electron beam in the toroidal cooler sections, the vertical position of the electron beam center changes according to 
\begin{equation}\label{eq:y}
y(z) \approx \int_{0}^{z}\theta(z')\,dz' = A_1 y_1(z) + A_2 y_2(z),
\end{equation}
where
\begin{equation}\label{eq:yi}
y_i(z) = z + k_i\ln\left[\frac{1+\exp\left(\frac{\displaystyle z_i-z}{\displaystyle k_i}\right)}{1+\exp\left(\frac{\displaystyle z_i}{\displaystyle k_i}\right)}\right]
\end{equation}
for $i=1,2$. When $y(z)$ becomes larger than $r_e=\chi r_c$ the overlap between electron beam and ion beam becomes zero (neglecting the ion beam radius). For $\chi =100$ the loss of beam overlap occurs at $z^\mathrm{(max)} = 646$~mm (figure~\ref{fig:angle}), which represents the maximum $z$ that has to be considered in the Monte-Carlo convolution. To account for the specifics of the CRYRING@ESR cooler the Monte-Carlo code from \cite{Wang2024} has been adapted accordingly.

\section{Results for electron-ion collision spectroscopy of Pb$^{78+}$}

\begin{figure}
	\includegraphics[width=\linewidth]{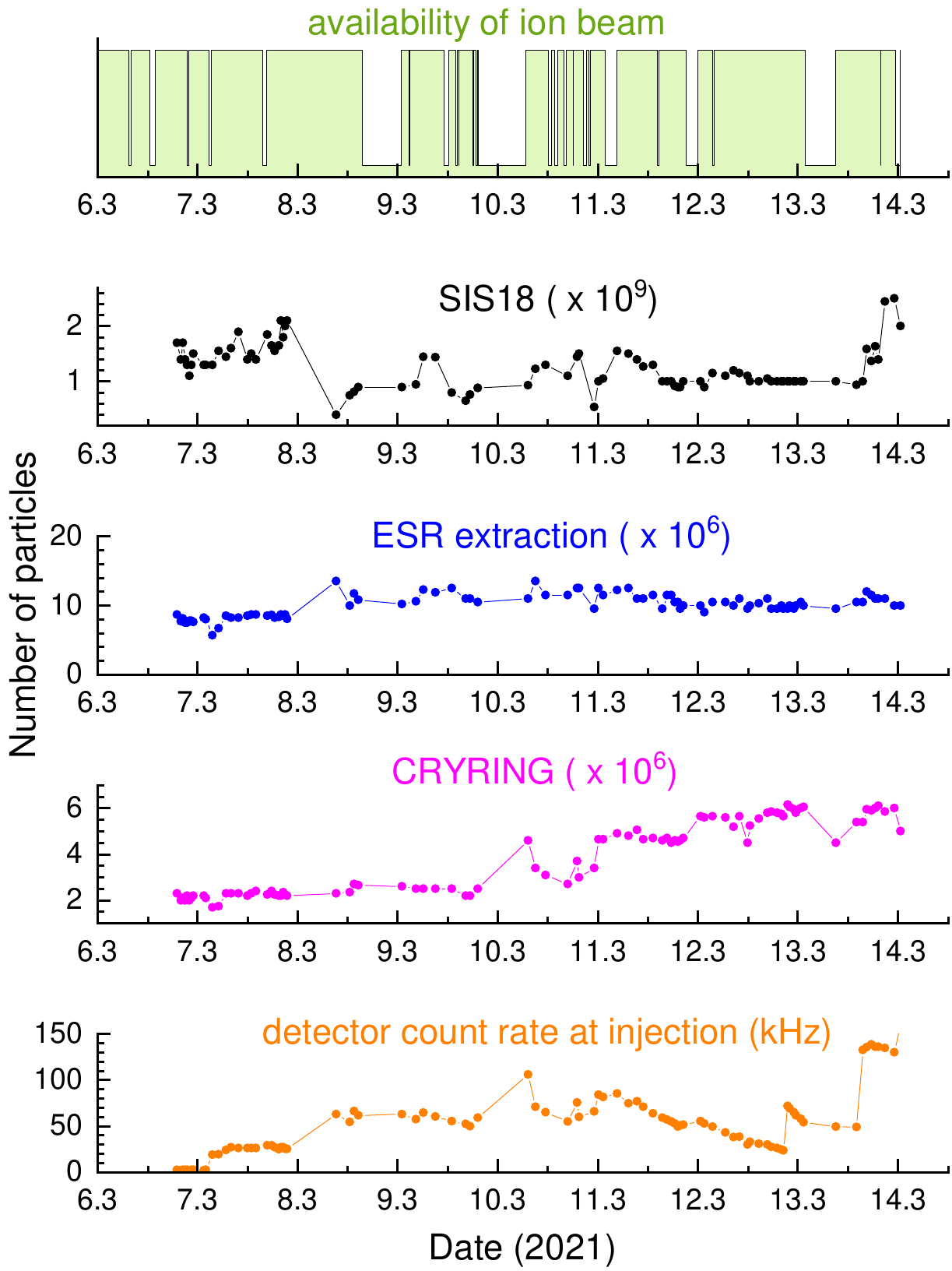}
	\caption{\label{fig:E131}Beam availability, particle numbers and detector counts in the course of beamtime E131 in March 2021.}
\end{figure}

The first beamtime at CRYRING@ESR with ions from the GSI accelerator chain was on electron-ion collision spectroscopy of Pb$^{78+}$ ions. It took place in March 2021 after a longer shutdown of the accelerator facility, during which the hardware and software of the accelerator control system had been entirely renewed. Remarkably, the new control system worked out of the box. Figure \ref{fig:E131} presents some key parameters of the beamtime.

Lead ions were first accelerated by UNILAC and injected into SIS18, which further accelerated them to an energy of 54~MeV/nucleon. The desired charge state $q=78$ was obtained by passing the ions through a stripper foil located in the extraction beam line of SIS18. On the average, $10^9$ ions per pulse were subsequently injected into ESR. There, the ions were cooled and decelerated to the final ion energy of 11.3~MeV/nucleon by operating the ESR in synchrotron mode. Eventually, up to $6\cdot10^6$ $^{208}$Pb$^{78+}$ ions were injected into CRYRING@ESR. The storage lifetime of the ion beam in CRYRING@ESR was about 30~s. The detector count rate followed the number of injected ions for four days. Later the detection efficiency deteriorated. This was unnoticed for some time and later remedied by increasing the detector voltages. The variation of the detection efficiency in the course of the beamtime was properly accounted for in the data analysis.

\begin{figure}
	\includegraphics[width=\linewidth]{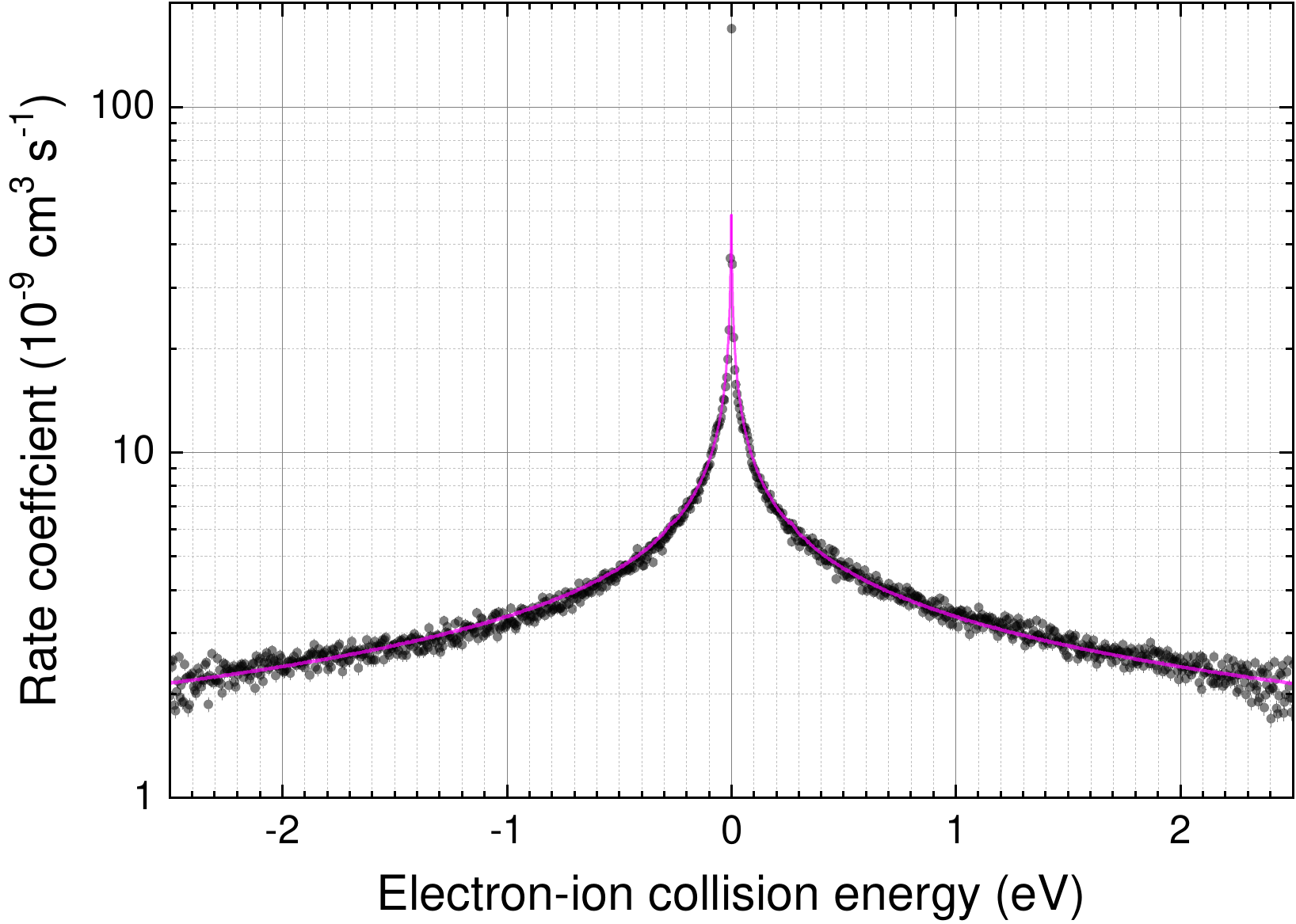}
	\caption{\label{fig:RR}Measured (symbols) and calculated (full line) merged-beams rate coefficient for electron-ion recombination of Pb$^{78+}$ ions in the vicinity of the RR peak at $E_\mathrm{cm}=0$. At positive (negative) energies the electrons have been faster (slower) than the ions. }
\end{figure}

Figure~\ref{fig:RR} shows the measured rate coefficient around zero electron-ion collision energy. Negative and positive energies indicate that the electrons have been slower and faster than the ions, respectively. The  symmetry of the experimental rate coefficient around $E_\mathrm{cm}=0$~eV can be exploited for an accurate determination of the cooling energy. For the present experiment this yields $E_\mathrm{cool}=6197.9\pm0.2$~eV (see \cite{Schippers2025} for details).

The measured rate coefficient in figure~\ref{fig:RR} peaks at $E_\mathrm{cm}=0$~eV because the cross section for nonresonant radiative recombination (RR) diverges at zero energy. The full line in figure~\ref{fig:RR} is the theoretical result for RR of Pb$^{78+}$ where a semiclassical hydrogenic approximation \cite{Bethe1957,Schippers2001c} has been used for the RR cross section in the  Monte-Carlo convolution. This approximation has been shown to be appropriate even for RR of U$^{92+}$ \cite{Shi2001b}, except for very low energies $E_\mathrm{cm}\lesssim k_BT_\perp$, where the experimental rate coefficient is larger than the theoretical one by in the present case a factor of 3.5 at $E_\mathrm{cm} = 0$~eV. This so called \lq{}recombination rate enhancement\rq\ has been observed previously at CRYRING \cite{Gao1995} and other storage rings \cite{Gwinner2000,Shi2001b,Hoffknecht2001b} and found to be an artifact of the merged-beams arrangement at electron coolers. This intriguing phenomenon became an attractive topic of research in itself \cite{Spreiter2000,Hoerndl2005c,Banas2009,Shi2025}.

\begin{figure}
	\includegraphics[width=0.96\linewidth]{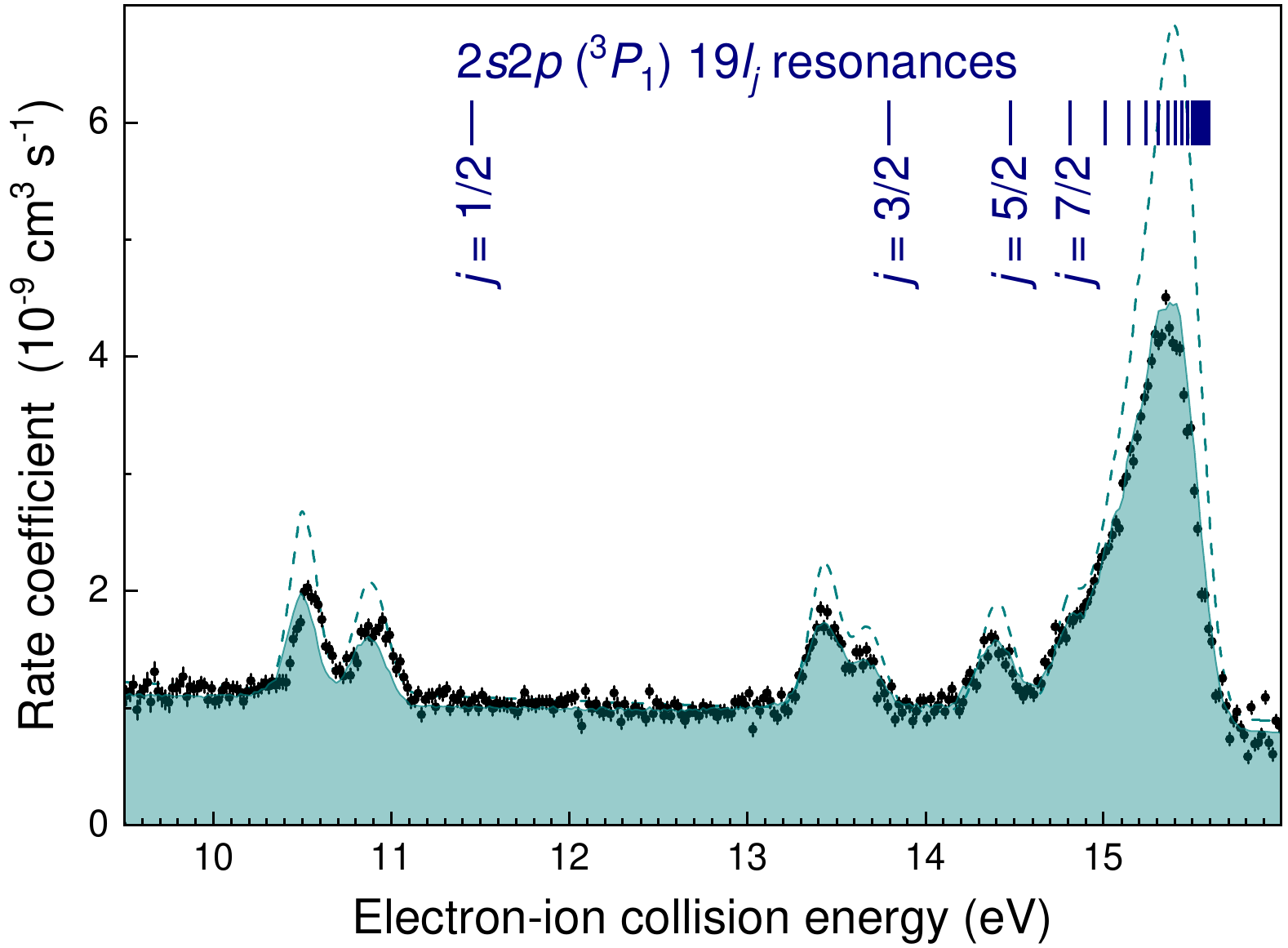}
	\caption{\label{fig:DR}Measured (symbols) and calculated (shaded curve) merged-beams rate coefficient for electron-ion recombination of Pb$^{78+}$ ions in the energy range of the $2s\,2p\,(^3P_1)\,19\ell_j$ resonances associated with $2s\to2p$ core excitations. The theoretical cross sections were obtained from the JAC atomic structure code \cite{Fritzsche2025} and convolved by the Monte-Carlo code discussed in section \ref{sec:mc}. {\color{black}The dashed line results from a convolution which disregards the toroidal effects in the electron cooler.} For the comparison an overall shift of 0.34~eV was applied to all theoretical resonance positions. The vertical bars mark the positions of the individual $j$ resonances as obtained from the Dirac formula for hydrogenic ions for the $19\ell_j$ Rydberg binding energies (see text).}
\end{figure}

Results for resonant recombination of Pb$^{78+}$ are presented in figure~\ref{fig:DR}. The resonances are associ\-ated with a resonant $2s\to2p$ excitation of the berylliumlike core and a simultaneous capture of the initially free electron into the $n=19$ Rydberg shell. The designation of the resonances is $2s\,2p\,(^3P_1)19\ell_j$. Their resonance energies  $E_\mathrm{res} = E_\mathrm{exci}-E_b(n\ell_j)$ are differences between  the core-excitation energy $E_\mathrm{exci}$ and the binding energies $E_b(19\ell_j)$ of the $19\ell_j$ Rydberg electron. Approximative values for $E_\mathrm{res}$ are obtained by using the hydrogenic Dirac formula for $E_b(19\ell_j)$ (see \cite{Bernhardt2015a} for details) and $E_\mathrm{exci}=244.937$~eV~\cite{Schippers2025}. The agreement with the measured resonance positions is not perfect, but sufficient for their unambiguous assignment. The splittings of the $j=1/2$ and $j=3/2$ resonances is due to the interaction between the excited berylliumlike core and the Rydberg electron. At higher $j$ values the splitting is too small to be discernible at the given experimental resolving power, which does not allow for an individual resolution of the $j\geq 7/2$ levels, either.

The shaded curve in figure~\ref{fig:DR} is the result of a fully relativistic calculation with the JAC atomic code \cite{Fritzsche2025}. The theoretical cross section has been convolved with the above mentioned Monte-Carlo code (section~\ref{sec:mc}). In the convolution the electron beam temperatures $kT_\| = 0.23$~meV and $kT_\perp = 3.3$~meV have been used. The latter value corresponds to what is expected from the experimental electron beam expansion factor $\chi = 33$. The parallel temperature is a factor of about 2 larger than what was found in earlier measurements at the CRYRING cooler. One of the reasons for this discrepancy is probably a larger-than-expected slew rate which is exhibited by the electron beam acceleration voltage upon voltage jumps of the HV amplifier \cite{Koutsostathis2024}.

{\color{black}Apart from an overall energy shift of 0.34~eV, the agreement between experiment and JAC calculation is excellent if the toroid effects in the electron cooler (figure~\ref{fig:angle}) are accounted for in the Monte-Carlo convolution of the theoretical cross section (shaded curve in figure~\ref{fig:DR}). The nonzero angles $\theta>0$ in the toroidal cooler sections lead to electron-ion collision energies (see equations~\ref{eq:gammarel} and~\ref{eq:Ecm}) that deviate from the nominal one realized in the straight electron-ion interaction zone (figure~\ref{fig:setup}) where $\theta=0$. This results in a smearing out of DR resonance strengths over a wide range of energies beyond the energy range of figure~\ref{fig:DR}. Conversely, there are significant differences between experiment and theory when the toroid effects are neglected  (dashed curve in figure~\ref{fig:DR}). 

The above mentioned 0.34-eV energy shift is required because decisive QED contributions are not (yet) treated in the JAC calculations \cite{Fritzsche2025}. Second order QED contributions to electron binding energies in berylliumlike ions have been evaluated only recently \cite{Malyshev2024}. When these are accounted for there is excellent agreement between theoretical and experimental Pb$^{78+}$ recombination resonance positions. A thorough discussion on how the experiment constrains the second-order QED calculations is out of the scope of the present paper. It can be found in another, more focussed publication \cite{Schippers2025}.}

\section{Conclusions and Outlook}\label{sec:Conclusions}

Electron-ion collision spectroscopy could be successfully established at CRYRING@ESR. Even in the very first experiment the resonance positions could be determined to within 30 meV uncertainty (as shown in \cite{Schippers2025}), which is less than the uncertainty of very recent associated state-of-the-art calculations employing strong-field QED to second order \cite{Malyshev2024}. 

Since the Pb$^{78+}$ beamtime, much effort has been taken to improve the experimental conditions for electron-ion collision spectroscopy at CRYRING@ESR. Among the measures taken were the provision of more accurate high-voltage probes (cf.\ section \ref{sec:HV}; in March 2021 only the Arnold HV probe was available). In addition, a new drift tube inside the electron cooler \cite{Koutsostathis2024} should allow for faster voltage jumps compared with the variation of the potential of the full cooler terminal with its much higher capacity.  Meanwhile, the CRYRING@ESR vacuum could also be considerably improved. In future experiments, this will lead to less background from ion collisions with residual-gas particles and to longer beam storage times.

We expect that in future experiments, the uncertainty of the experimental energy scale can be reduced by an order of magnitude as compared to the Pb$^{78+}$ experiment. Data from a follow-up beamtime in March 2024 with berylliumlike Au$^{75+}$ ions are currently being analyzed. A new experiment  proposal for an experiment with berylliumlike U$^{88+}$ ions has recently been submitted to the GSI Program Advisory Committee. Theoretical calculations suggest that this ion supports strong recombination resonances in the 1--2~eV collision-energy range, where the experimental energy uncertainty is estimated to be at the few-meV level (figure~\ref{fig:error}). This would allow one to challenge strong-field QED calculations beyond the second order.

\section*{Acknowledgments}

The results presented here are based on the experiment E131, which was performed at the heavy-ion storage ring CRYING@ESR at the GSI Helmholtzzentrum fuer Schwerionenforschung, Darmstadt (Germany) in the frame of FAIR Phase-0. Financial support from the German Federal Ministry for Education and Research (Bundesministerium f\"ur Bildung und Forschung, BMBF) via the Collaborative Research Center ErUM-FSP T05 --- \lq\lq Aufbau von APPA bei FAIR\rq\rq\ (Grant Nos.\ 05P19PMFA1, 05P19RGFA1, 05P21PMFA1, 05P21RGFA1, and 05P24RG2) is gratefully acknowledged. C.~B.\ and S.-X.~W.\ acknowledge the support by the State of Hesse within the Research Cluster ELEMENTS (Project ID 500/10.006).

\section*{References}

\providecommand{\newblock}{}

\end{document}